\documentstyle[11pt]{article}        
\input{psfig}

\begin{document}               
\title{Indian Physics: Outline of Early History}

\author{Subhash Kak}
\maketitle
\thispagestyle{empty}

\section{Introduction}

Historians of science are generally unaware of the contributions of Indians 
to physics. The main reason for this 
is that very little research has been done on the subject
since Seal's {\it The Positive Sciences of the Hindus} appeared in 1915,
a consequence of the fact that
there are few history of science departments in Indian universities.
Work relevant to the history
of Indian physics has been done in philosophy departments but
this is generally inaccessible to historians of physics.
Some other work concerning history of ideas in physics has been
published by historians of astronomy.

The objective of this paper is to present a preliminary outline of early history
of physics in India.
The focus here are the schools of Vai\'{s}e\d{s}ika and
S\={a}\d{m}khya that were interested in general principles of
atomic theory and cosmology.
Physical ideas in these and other schools were applied 
to technology as we can see for a much later period in 
Dharampal's book [4].

This paper should be read in conjunction with the papers on history 
of early Indian science and astronomy by the author [14-18, 22, 23],
where the relevant Vedic ideas on cosmology are presented. 
To summarize this background context, 
the Vedic texts present a tripartite and recursive world view.  The
universe is viewed as three regions of earth, space, and sky which
in the human being are mirrored in the physical body, the breath,
and mind. The processes in the sky, on
earth, and within the mind are taken to be connected.  The universe is
mirrored in the cognitive system, leading to the idea that introspection
can yield knowledge.

\section{On classficiation}

The Vedic seers speak of \d{r}ta, the laws
underlying the univerese. They also assert that
all descriptions are limited, and outside
their normal context they lead to logical paradox.  
The notable features of this world view are:

\begin{itemize}

\item
An infinitely old, cyclic universe 

\item
An atomic world and the subject/object dichotomy

\item
Relativity of time and space

\item
Evolution of life

\item
A science of mind

\item
Computable laws

\item
Language theory

\end{itemize}

As one would expect, these conceptions of evolution of life, relativity of space
and time, and science of mind are not quite on the same lines as
that of contemporary science. But that is precisely what makes 
Indian science especially interesting to the historian.

The first step in the development of any science is the
naming of objects and categories.
Then come the questions of change and transformation and
the recognition that a certain essence of reality
is unaffected by change.
Having named objects and events, one turns
to the relationships
between them.
The enumeration of categories in groups and their
relationship with other such groups comes
later. 
The question of the nature of the cognitive
process by which the knowledge of the universe is
obtained also comes at the end.

During the \d{R}gvedic period itself, it had come to be
recognized that although nature follows
laws, a certain freedom characterizes human behaviour.
The fundamental unity of reality is thus split into
two distinct categories related to innate nature and
cognition.
The universe not only exists outside of ourselves,
but a copy of it, howsoever imperfect, exists within
each one of us.
The enumeration of categories as they arise in the
space of the mind is the concern of S\={a}\d{m}khya.
The stated objective is to obtain 
discriminative knowledge of the manifest ({\it vyakta}),
the unmanifest ({\it avyakta}) and the knower
({\it puru\d{s}a}).\footnotemark
On the other hand, Vai\'{s}e\d{s}ika deals with
the goal attributed to Ka\d{n}\={a}da,
the mythical founder of the system,
{\it yad iha bh\={a}var\={u}pam, tat sarva\d{m}
may\={a}upasa\d{m}khy\={a}tavyam}, ``I shall enumerate
everything [in this world] that has the
character of being.''

The two systems have differing focus.
S\={a}\d{m}khya addresses evolution at the
cosmic and the psychological levels;
Vai\'{s}e\d{s}ika delves deeper into the
nature of substances and
its scope includes both
physics as well as metaphysics.

The emphasis in Indian thought on knowing the
outside through an analysis of cognitive categories was
far in advance of the concepts used by
historians of science until the rise of
modern physics.
As a result, the 
six {\it dar\'{s}anas} were
often misrepresented in the 
commentaries that were written with the
rise of Indian studies in the nineteenth century.
These mistakes have been repeated in more recent works
because this commentatorial tradition still 
operates within the framework of reductionist 
physics and analysis.
With the rise of relativity and quantum
mechanics, the subject has become central in
the understanding of the physical universe.
The outer world exists because there is someone to
perceive it; likewise the mind is characterized
by the associations between various objects and
processes of the outer world.\footnotemark
An examination of the physical world in terms of
categories of the mind or of ``being'' constitutes
a perfectly legitimate way of approaching the
outer world, albeit it is different from the manner
in which Western science developed.

S\={a}\d{m}khya and Vai\'{s}e\d{s}ika are generally
paired with Yoga and Ny\={a}ya, respectively.
The reason behind such a pairing is that
the paired system provides the student with the
ability to make further progress
in his understanding.
The focus in S\={a}\d{m}khya is the inner world
and, therefore, an experiential or meditative
attitude complements it. 
The insights of Yoga validate the
categories of S\={a}\d{m}khya, indeed the two
could proceed in a complementary fashion, which
is why the two are considered the
same system sometimes.
In Vai\'{s}e\d{s}ika, 
the focus is more on an enumeration of the categories of
being, perceived apart from oneself.
Since the categories are
very many,  the use of formal logic is essential
to draw inferences,
and in this respect
Ny\={a}ya is its sister system.

Actually, Ny\={a}ya (logic) provides the
analytical basis for all Indian sciences.
Naiy\={a}yikas say
{\it astitva j\~{n}eyatva abhidheyatva},
``whatever exists, is knowable and nameable.''
But it is also stated that speech has
four forms, of which one kind, the par\={a}, is unmanifest.\footnotemark  
So all description and analysis is ultimately limited by paradox.

The categories of 
S\={a}\d{m}khya and
Vai\'{s}e\d{s}ika 
describe
the physical and the psychological worlds.
A comprehensive theory,
integrating the insentient and the sentient, is offered.

We first provide a brief review of the 
the S\={a}\d{m}khya and the Vai\'{s}e\d{s}ika categories
and then examine the significance of their
physical concepts.
The beginnings of these concepts can be traced 
back to the Vedic literature.

\section{Overview and early development}
We first begin with a few remarks on the chronology of the Indian
texts.
New results in archaeology have shown that the Indian tradition can
be traced back in a series of unbroken phases to at least
8000 B.C.E.
Archaeologists and geologists also believe that 
the Sarasvat\={\i}, the preeminent river of the
\d{R}gvedic age, dried up around 1900 B.C.E.,
leading to the collapse of the towns of the Harappan era that
were primarily distributed in the Sarasvat\={\i} region.
There is increasing
acceptance of the view that the \d{R}gveda should be earlier than
1900 B.C.E.
The early Br\={a}hma\d{n}as and the Upani\d{s}ads then belong to the
second millennium B.C.E.\footnotemark

The Vedic hymns speak of ideas
that are later 
described at greater length in the
{\it dar\'{s}anas}.
Nature has an order that is 
expressed as {\it \d{r}ta}.\footnotemark 
This order is behind the regularity in the movements of the planets,
the seasons, and cycles on earth.
\d{R}ta defines an inflexible law of harmony
which offers a basis for its comprehension through
the mind.
The principle of order is sometimes represented by
the pillar ({\it skambha}) as in the Atharvaveda\footnotemark and 
anthropomorphized as
Brahma\d{n}aspati.\footnotemark

The \d{R}gvedic hymn 10.129 describes how prior to a separation between
the subject and the object neither space or time existed.
It goes on to say:

\begin{quote}
In the beginning desire arose, born of the mind, it was the primal seed.
The seers who have searched their hearts with wisdom know the
connection ({\it bandhu}) between being and non-being.\\

A cord stretched across them; what was above, and what was
below? Seminal powers made mighty forces, below was strength
and above was impulse.\footnotemark

\end{quote}

The connections ({\it bandhu}) between the
outer and the inner are affirmed.
Next, there is mention of the dichotomy 
between puru\d{s}a and prk\d{r}ti, the impulse and
the strength.

In \d{R}gveda 10.90, puru\d{s}a, is
the cosmic person out of whose dismembered body the
living and the inanimate worlds emerge.
Here too a dichotomy, expressed through the
symbols of male and female, marks the paradoxical
beginning of empirical existence.
Puru\d{s}a is born out of {\it vir\={a}j}, ``the shining one,'' and
she out of him.
This marks a distinction between puru\d{s}a as
transcendent reality and its manifestation in terms
of individual consciousness.

Further on in the same hymn, several categories related
to existence, such as space, sky, earth, directions, 
wind, metres and so on are created.
Such an enumeration is described at greater length in
the dialogue in the
B\d{r}had\={a}ra\d{n}yaka Upani\d{s}ad between Y\={a}j\~{n}avalkya and Maitrey\={\i}
where seventeen of the twenty-three categories of
classical S\={a}\d{m}khya are noted:

\begin{quote}
As all waters find their goal in the sea, so all touches
in the skin, all smells in the
nose, all taste in the tongue,
all forms in the eye, all sounds in the ear, all
deliberations in the mind, all knowledge in the intellect, all actions
in the hands, all enjoyment in sex,
all elimination in the excretory organs, 
all movement in the feet, and all the Vedas in
speech.\\

As a mass of salt has neither inside nor outside, but is
altogether a mass of taste, thus indeed has that Self
neither inside nor outside, but is altogether a mass
of knowledge; and having risen from these elements,
vanishes again in them.\footnotemark

\end{quote}

These include the five material elements, the five
organs of sense, the five organs of action, 
the {\it buddhi}, in the form of {\it vij\~{n}\={a}na},
{\it aha\d{m}k\={a}ra},\footnotemark and mind.
The only categories of the
late S\={a}\d{m}khya which are not explicitly mentioned
in the 
B\d{r}had\={a}ra\d{n}yaka Upani\d{s}ad 
are the {\it tanm\={a}tras}, but the
{\it bandhu} between the gross and the subtle,
which is emphasized again and again in the \d{R}gveda,
indicates the implicit recognition of the corresponding subtle
{\it tanm\={a}tra} for the five gross elements.
This subtle representation of the outer in
terms of {\it m\={a}tr\={a}} is
described explicitly
in the Kau\d{s}\={\i}taki Br\={a}hma\d{n}a Upani\d{s}ad
where the specific abstract correspondences for certain
outer functions, such as speech, breath, order, and so on, are listed in terms of 
{\it bh\={u}tam\={a}tr\={a}}.\footnotemark
The word {\it m\={a}tr\={a}} here refers to the essence in
the same manner as 
in the notion of tanm\={a}tra.
Aha\d{m}k\={a}ra is described in the Chandogya Upani\d{s}ad
as the one who sees the universe.\footnotemark

In other words, all the elements of S\={a}\d{m}khya seem to
be in place in the Vedic literature.
We also have a proper scientific system with its
cosmic order and 
corresponding laws (\d{r}ta), entities and relationships.
Even the workings of the human mind are
subjected to logical analysis.

The Vedic system is a tripartite
and recursive world view.
At the most basic level,
the universe is viewed as three regions of earth, space, and sky
with the corresponding entities of Agni, 
Indra, and Vi\'{s}ve Deva\d{h} (all gods).
These three regions are represented in the Vedic ritual as three
different altars.
There is a mapping of these regions within the human body as well.
The Ch\={a}ndogya Upani\d{s}ad speaks of a tripartite 
manifestation of reality, expressed as fire (red),
water (white), and food (black) correlated with
speech, breath, and mind.\footnotemark
\'{S}vet\={a}\'{s}vatara Upani\d{s}ad also describes the
red, white, and black aspects of the One.\footnotemark
In \={A}yurveda, the three {\it do\d{s}as} (humours), 
{\it v\={a}ta, pitta}, and {\it kapha}, likewise define a
tripartite model.

Counting separately the joining regions leads to a total
of five categories where,
as we see in Figure 1, water separates earth and fire, and
air separates fire and ether.\footnotemark
This counting in groups of five is seen in a variety of
contexts as in the five directions,
five senses, five seasons, five metres, five chants,
five peoples, five breaths, and so on.

\vspace{8mm}
\begin{figure}
\hspace*{0.2in}\centering{
\psfig{file=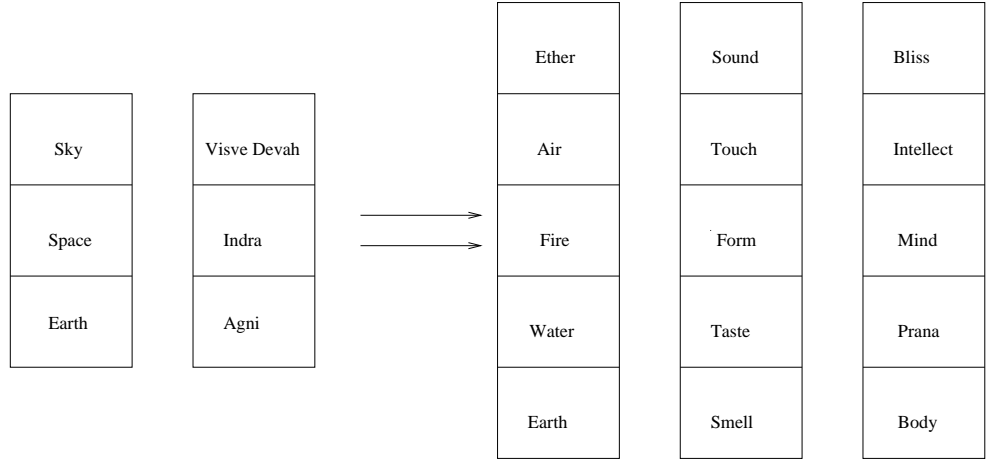,width=13cm}}
\caption{Five-layered universe}
\end{figure}
\vspace{8mm}

Although
the processes in the sky, on earth, and within the mind are connected,
all descriptions of the universe lead to logical paradox.
The one category transcending all oppositions is
{\it brahman}.
Vedic ritual is a symbolic representation of this world view.

The 
complementarity between
the mind and the outer world is of
fundamental significance.
Knowledge is classified in two ways: the lower ({\it apar\={a}}) or dual; and the
higher ({\it par\={a}}) or unified.
Knowledge is superficially dual and
paradoxical but at a deeper level it has a unity.
The material and the conscious
are aspects of the same transcendental reality.

In the Ch\={a}ndogya Upani\d{s}ad,
Udd\={a}laka \={A}ru\d{n}i describes the unity behind the
apparent duality as {\it sadvidy\={a}}.
Being ({\it sat}) provides both the origin and the unity:

\begin{quote}
In the beginning, my dear, this world was just Being, one only,
without a second.
Others say: ``In the beginning this world was just Nonbeing ({\it asat}),
one only, without a second; from that Nonbeing Being was produced.''
But, my dear, how could this be?
How from Nonbeing could Being be produced? 
No, my dear, in the beginning this world was just Being,
one only, without a second.\footnotemark 
\end{quote}
 
In the Taittir\={\i}ya Upani\d{s}ad, the
individual is represented in terms of five different
sheaths or levels that enclose the individual's self.\footnotemark
This represents another instance of expanded
tripartite model.
These levels, shown in an ascending order, are:
\begin{itemize}
\item The physical body ({\it annamaya ko\'{s}a})
\item The energy sheath ({\it pr\={a}\d{n}amaya ko\'{s}a})
\item The mental sheath ({\it manomaya ko\'{s}a})
\item The intellect sheath ({\it vij\~{n}\={a}namaya ko\'{s}a})
\item The bliss sheath ({\it \={a}nandamaya ko\'{s}a})
\end{itemize}
 
These sheaths are defined at increasingly finer levels.
At the highest level, above the bliss sheath, is the self.
Intellect is placed below bliss,
which is a recognition of the fact that eventually
meaning is communicated not by associations, but rather by
a synthesizing vision expressed by the notion of bliss.
 
Pr\={a}\d{n}a is the energy
coursing through
the physical and mental processes.
If one looked at the individual in the three fundamental levels, then
at the lowest level is the physical body, at the next higher level
is the energy systems at work, and at the next higher level are
the thoughts.
Since the three levels are interrelated, the energy situation may
be changed by inputs either at the physical level or at the
mental level.

The key notion is that each higher level represents characteristics that
are emergent on the ground of the previous level.
In this theory mind is an emergent entity, but this emergence requires
the presence of the self.
 
The mind may be viewed in a five-fold way:
manas, aha\d{m}k\={a}ra, citta, buddhi, and \={a}tman.
Again these categories parallel those of Figure 1.

The notions of enumeration and
indivisibility
are so pervasive in Vedic thought that it is impossible
to put a date on the rise of
S\={a}\d{m}khya and 
Vai\'{s}e\d{s}ika.
But there developed specific schools where a
particular manner of defining the attributes was
taken;
these schools trace their
lineage to specific individuals, often starting with a
mythical \d{r}\d{s}i.

\subsection*{Classical S\={a}\d{m}khya}

The notions of S\={a}\d{m}khya form
a part of the earliest Vedic texts.
As a system called by its formal name,
it is described in the Mok\d{s}adharma and the
Bhagavad G\={\i}t\={a} as well as in the
Upani\d{s}ads.
Its legendary founder was the sage Kapila
who used to be dated to around 7th century B.C.E., but in light
of the new findings related to Indian antiquity, is likely to have
lived much earlier than that.
The texts speak of at least twenty-six teachers
including \={A}suri, Pa\~{n}ca\'{s}ikha, Vindhyav\={a}sa,
V\={a}r\d{s}aga\d{n}ya, Jaig\={\i}\d{s}avya,
and 
\={I}\'{s}varak\d{r}\d{s}\d{n}a.
By ``classical S\={a}\d{m}khya'' we mean the {\it S\={a}\d{m}khya-K\={a}rik\={a}} (SK)
of \={I}\'{s}varak\d{r}\d{s}\d{n}a.
The S\={a}\d{m}khya-K\={a}rik\={a}
claims to be the summary of an earlier, more comprehensive
treatise, the {\it \d{S}a\d{s}\d{t}itantra}.

According to S\={a}\d{m}khya, reality is composed of a number of
basic
principles ({\it tattva}), which are taken to be 
twenty-five in the classical 
system. But since the heart of the system is its
hierarchical framework, the exact number of the principles 
varies, especially in the earliest writings.
But such a variation is of no fundamental importance.

In the classical system, 
the first principle is (1) {\it prak\d{r}ti}, which is taken to
be the cause of 
evolution.
From prak\d{r}ti develops (2) intelligence ({\it buddhi}, also called
{\it mahat}), and thereafter (3) self-consciousness ({\it aha\d{m}k\={a}ra}).
From self-consciousness emerge the five subtle elements
({\it tanm\={a}tra}): (4) ether ({\it \={a}k\={a}\'{s}a}),
(5) air, (6) light, (7) water, and (8) earth.
From the subtle elements emerge the five
(9-13) material elements ({\it mah\={a}bh\={u}ta}).
Next emerge the five organs of sense ({\it j\~{n}\={a}nendriya}):
(14) hearing, (15) touch, (16) sight, (17) taste, and
(18) smell, and five organs of action ({\it karmendriya}):
(19) speech, (20) grasping, (21) walking, (22) evacuation, and
(23) procreation.

Finally, self-consciousness produces the twenty-fourth of the
basic elements: (24) mind ({\it manas}), which, as a sixth sense,
mediates between the ten organs and the outside world.
The last, twenty-fifth, tattva is (25) {\it puru\d{s}a.}

\vspace{8mm}
\begin{figure}
\hspace*{0.2in}\centering{
\psfig{file=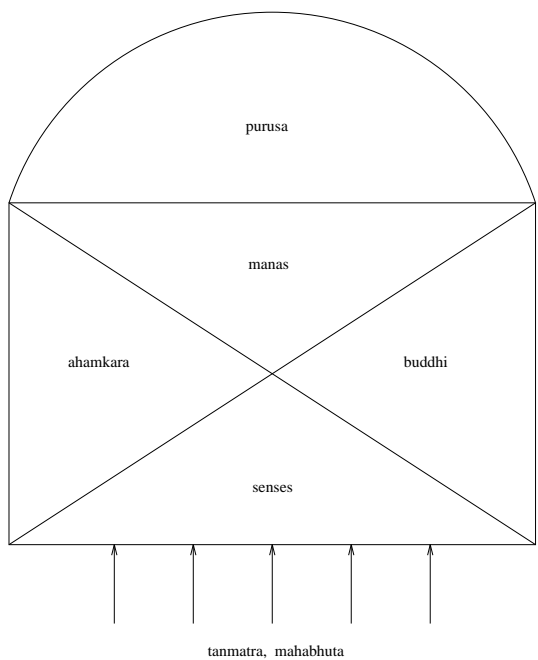,width=7cm}}
\caption{The cognitive system}
\end{figure}

\vspace{8mm}

The emergence from prak\d{r}ti of intelligence and, later,
of subtle and gross elements, mind and consciousness, appears
to mirror the stages through which a newly-conceived individual
will pass.
Here intelligence, as the second tattva, is what endows
the newly fertilized cell the ability to organize and grow;
self-consciousness represents the stage which allows the
organism to sense the environment, and so on.
The thesis that the world is connected, allows 
one to see the same process at the cosmic and the psychological
levels.

The doctrine of the three constituent qualities ({\it gu\d{n}a}):
{\it sattva, rajas}, and {\it tamas}, plays a very important role in
the S\={a}\d{m}khya physics and metaphysics.
These gu\d{n}as are described in the Upani\d{s}ads. 
In its undeveloped state, cosmic matter has these 
{\it gu\d{n}as} in equilibrium.
As the world evolves, one or the other of these become
preponderant in different objects or beings,
giving specific character to each.
The quality of sattva, which stands for virtue or
transparence, inheres in all things tending to truth,
wisdom, beauty or goodness;
the quality of rajas, or activity, energy or passion, is present
in all that is fierce, forceful or active;
the quality of tamas, which stands for inertia,
is to be found in all that is stupid or dull.
The gu\d{n}as can be viewed as the three constituent
strands of materiality.

S\={a}\d{m}khya can also be seen as having three basic dimensions:

\begin{enumerate}

\item The constitutive (tattva) dimension, dealing with form ({\it r\={u}pa}),
the principle or the essential core ({\it li\.{n}ga});

\item The projective ({\it bh\={a}va}) dimension, concerning the projective or
the intentional ({\it prav\d{r}tti}),
the predispositional, or cause-effect
({\it naimittanaimittika}); and

\item The consequent ({\it phala}) dimension, dealing with what has
come to pass ({\it bh\={u}ta}) or the phenomenal creation
({\it pratyayasarga}).
\end{enumerate}

They gu\d{n}as can also be viewed as the threads that tie together the three
realms of the tattvas, the bh\={a}vas, and the bh\={u}tas.

\subsection*{Vai\'{s}e\d{s}ika}
This school of ``individual characteristics'' is supposed
to have been founded by Ka\d{n}\={a}da, the son of
Ul\={u}ka.
Other important sages associated with this tradition include
Candramati, Pra\'{s}astap\={a}da, Vyoma\'{s}iva and Udayana.
Ka\d{n}\={a}da's
Vai\'{s}e\d{s}ika S\={u}tras (VS)
describe a system of
physics and metaphysics.
Its physics is an atomic theory of nature, where the atoms are
distinct from the soul, of which they are the instruments.
Each element has individual characteristics ({\it vi\'{s}e\d{s}as}),
which distinguish it from the other non-atomic
substances ({\it dravyas}): time, space, soul, and mind.
The atoms are considered to be eternal.

There are six fundamental categories 
({\it pad\={a}rtha}) associated with reality:
substance ({\it dravya}), quality ({\it gu\d{n}a}),
motion ({\it karman}), universal ({\it s\={a}m\={a}nya}),
particularity ({\it vi\'{s}e\d{s}a}), and inherence
({\it samav\={a}ya}).
The first three of these have a real objective 
existence and the last three are products of
intellectual discrimination. 
Each of these categories is further subdivided as follows.

There are nine classes of substances ({\it dravya}),
some of which are nonatomic, some atomic, and
others all-pervasive.
The nonatomic ground is provided by the three substances
ether ({\it \={a}k\={a}\'{s}a}),
space ({\it di\'{s}}), and time ({\it k\={a}la}),
which are unitary and indestructible;
a further four,
earth ({\it p\d{r}thiv\={\i}}), water ({\it \={a}pas}), fire
({\it tejas}), and air ({\it v\={a}yu}) are
atomic composed of indivisible, and indestructible atoms
({\it a\d{n}u, param\={a}\d{n}u});
self ({\it \={a}tman}), which is the eighth, is omnipresent and eternal;
and, lastly, the ninth, is the mind
({\it manas}), which is also eternal but of atomic
dimensions, that is, infinitely small.

\vspace{8mm}
\begin{figure}
\hspace*{0.2in}\centering{
\psfig{file=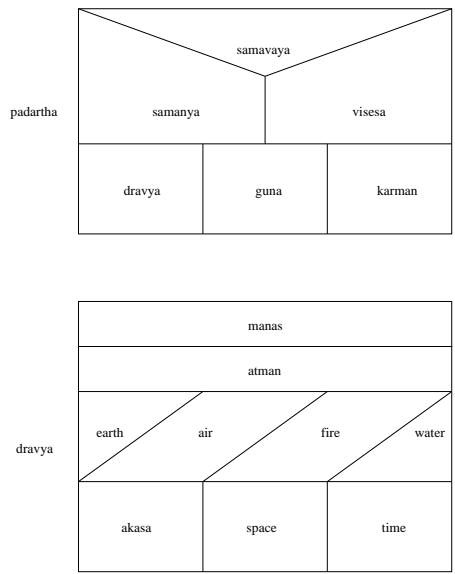,width=7cm}}
\caption{The categories of Vaisesika}
\end{figure}

\vspace{8mm}

There are seventeen qualities 
({\it gu\d{n}a}),
listed in no particular order as colour or form ({\it r\={u}pa}), taste
({\it rasa}), smell ({\it gandha}), and touch ({\it spar\'{s}a});
number ({\it sa\d{m}khy\={a}}), size or dimension
({\it parim\={a}\d{n}a}), separateness ({\it p\d{r}thaktva}), conjunction
({\it sa\d{m}yoga}), and disjunction ({\it vibh\={a}ga});
remoteness ({\it paratva}) and
nearness ({\it aparatva});  judgment ({\it buddhi}), pleasure ({\it sukha}), pain
({\it du\d{h}kha}), desire ({\it icch\={a}}),
aversion ({\it dve\d{s}a}), and effort ({\it prayatna}).
These qualities are either physical or psychological.
Remoteness and nearness are interpreted in two different ways:
temporally or spatially.
This list is not taken to be comprehensive because later
sound is also described as a quality.
But there is a fundamental difference 
between sound and light.
Sound is carried by the non-atomic \={a}k\={a}\'{s}a,
whereas light,
implied by r\={u}pa,
is carried by tejas atoms.
But even sound is sometimes seen as a specific characteristic of atoms.

There are five different types of motion 
({\it karman}) that are associated
with material particles or the organs of the mind:
ejection, falling (attraction), contraction, expansion, and composite
motion.

Universals ({\it s\={a}m\={a}nya})
are recurrent generic properties in substances,
qualities, and motions.
Particularities ({\it vi\'{s}e\d{s}a}) reside exclusively in the eternal,
non-composite substances, that is, in the
individual atoms, souls, and minds, and in the
unitary substances ether, space, and time.

Inherence ({\it samav\={a}ya}) is the relationship
between entities that occur at the same time.
This provides the binding that we see in the various
categories so that we are able to synthesize our
experience.

The Vai\'{s}e\d{s}ika atomic structure characterizes four
of the five S\={a}\d{m}khyan mah\={a}bh\={u}tas;
the fifth, ether, is non-atomic and
all-pervasive.
Some of the 
Vai\'{s}e\d{s}ika gu\d{n}as correspond to the
S\={a}\d{m}khyan tanm\={a}tras.
In S\={a}\d{m}khya the tanm\={a}tras come first, in
Vai\'{s}e\d{s}ika atoms are primary.

Each of the two schools has had a very long history.\footnotemark
This included many variations to the classical formulation given above.
There has also been considerable difference in interpretation.
In the sections that follow, I present an eclectic summary from this mass of material
to communicate their main physical ideas.

\section{Physical concepts}

The Vai\'{s}e\d{s}ika categories
appear to provide a convenient starting
point to examine the physical concepts inherent
in these two systems.

The ground layer consists of indivisible,
invisible and indestructible atoms
({\it a\d{n}u, param\={a}\d{n}u}).
It is the aggregation of these atoms that
give rise to different destructible
compound substances.
These atoms are ideals, representing unities
of fundamental attributes.
In this sense, they are quite similar to the
concept of such elementary particles of modern physics
which are proposed on theoretical grounds.

It is useful to consider the modern atomic doctrine for the
sake of reference.
Here the elementary particles are characterized by various
attributes, each of which has a numerical value.
These attributes include mass, charge, angular momentum,
energy, and so on.
The properties of bulk matter is, in principle, obtainable
from those of its constituents, but at each higher level
of aggregation of atoms, new properties emerge.

Philosophically, there are two main approaches,
{\it positivism} and {\it realism}, for the understanding
of physics.
According to the positivist,
the only scientific knowledge is the one that
can be expressed in logical statements.
Since our logic and our language is a result of the observations
of the world, this presupposes that the 
observer is central to this knowledge.
This is essentially the same
as the Ny\={a}ya position.
The realist believes that there exists an independent
reality which is probed through observation and
experiment.
Put differently, the positivists believe that knowledge
is subjective, whereas realists believe that it is objective.

A positivist accepts that there are elements of an
empirical reality which science uncovers, but points out
that the realist view involves a logical contradiction,
since there is no way of observing an observer-independent
reality and hence we cannot verify that such a reality exists.

A weaker form of objectivity is sometimes identified with
the positivist position.
Here we speak of an empirical reality which is not 
independent of the observer, but is the same for all
observers.
Such weak objectivity characterizes relativity theory.

\subsection*{Atoms and their combinations}

According to the Vai\'{s}e\d{s}ika 
S\={u}tras, ``Earth possesses colour, taste, smell, and
touch. Waters possess colour, taste, and touch, and are fluid and
viscid. Fire possesses colour and touch. Air possesses touch.
These (preceding characteristics) are not in ether.''\footnotemark
This indicates how the qualities are seen as being built out
of elementary entities.
Such a unitary picture is even more clearly spelt out for the atoms
and the tanm\={a}tras.
As mentioned before, S\={a}\d{m}khya provides a slightly
different focus, where the abstract tanm\={a}tras are considered to
be the building blocks for the gross atoms.

The Vai\'{s}e\d{s}ika atomic substances are defined in 
a matrix 
of four non-atomic substances ({\it dravyas})---time, space,
soul and mind.
In other words, the physical universe has an objective
existence and mind and soul do not simply
emerge from the material ground and disappear when
the material structure disintegrates.

The objective elements of the physical world
are characterized by 
dravya, gu\d{n}a, and karman, or substance,
quality, and action.
There is a further
characterization 
in terms of
non-reactive and reactive properties.

Two atoms combine to form a
binary molecule ({\it dvya\d{n}uka}).
Two, three, four or more dvya\d{n}ukas combine into
grosser molecules of trya\d{n}uka, catura\d{n}uka, and so on.
The other view is that
atoms form dyads and triads directly to form molecules
fo different substances.
Atoms possess an incessant vibratory motion.
The activity of the atoms and their combinations are not
arbitrary but according to laws that are expressed as
the ad\d{r}\d{s}\d{t}a.

Molecules can also break up under the influence of
heat ({\it p\={a}kajotpatti}).
In this doctrine of {\it p\={\i}lup\={a}ka} (heating of atoms),
the impact of heat particles decomposes a molecule.

Heat and light rays are taken to consist of very
small particles of high velocity.
Being particles, their velocity is finite.
This is also apparent from the fact that motion is
contingent upon time as
one of the dravyas.
The particles of heat and light
can be endowed with different characteristics
and so heat and light can be of different kinds.

Elsewhere it is said that there is no difference between the
atom of a barley seed and paddy seed, since these are but atoms
of earth.
Under the impact of heat particles, atoms can exhibit
new characteristics.

A bh\={u}ta-atom evolves out of integration from the corresponding
tanm\={a}tra.
This indicates a primacy of the abstract over the material.
On the other hand,
the atoms may be taken to be unitary objects and their combinations 
seen as generating various tanm\={a}tras.
One may further assume that rudiment-matter ({\it bh\={u}t\={a}di}) leads
to its more specific forms.
Brajendranath Seal summarizes some views on the
relationship between atoms and tanm\={a}tras as follows:

\begin{quote}

\item The rudiment-matter (bh\={u}t\={a}di) acted on by
rajas (energy) produces the sound-potential (vibration-potential).

\item The vibration-potential, as a radicle, with accretion of
atoms, condensing and collocating, generates the
touch-potential which is impingent as well as vibratory.

\item The impact-potential, as a radicle, with a similar
accretion of
atoms generates the
heat-and-light-potential which radiates
light and heat in addition to being impingent as well as vibratory.

\item The light-and-heat-potential, as a radicle, with further
accretion of
atoms
generates the taste-potential.

\item The taste-potential, as a radicle, with further accretion of
atoms, generates the
smell-potential.\footnotemark
\end{quote}

The order of the formation of the bh\={u}ta-param\={a}\d{n}u
is seen according to the following hierarchical scheme:

\begin{enumerate}

\item The sound-potential, subtile matter, with 
accretion of rudiment-matter generates the
\={a}k\={a}\'{s}a atom.

\item The touch-potential combines with vibratory particles 
(sound-potential) to generate the
v\={a}yu atom.

\item The light-and-hear-potentials combine with touch and
sound-potentials to produce the
tejas atom.

\item The taste-potential combines with the foregoing three to produce
the
\={a}pas atom.

\item The smell-potential combines with the foregoing four to
generate the
earth atom.\footnotemark

\end{enumerate}

In summary, all these views see matter as being
of a unitary nature which when excited to different
states produces potential of different
kinds that correspond to the
tanm\={a}tras and then constitutes different
elements.

The {\it Pad\={a}rthadharmasa\d{m}graha} of 
Pra\'{s}astap\={a}da deals with the
question of ultimate substances.
Earth, fire, water, and earth are here taken to 
be the basic material substances.
But their existence is taken to be contingent on the 
presence of someone who knows of them, namely Brahman.
Pra\'{s}astap\={a}da's
commentary and exposition of the relevant s\={u}tras of
VS, with s\={u}tra numbers shown in parentheses, is as follows:

\begin{quote}
\={A}k\={a}\'{s}a (ether), time and space have no lower
constituents. (VS 2.1.27, 29-31)\\
Of 
\={a}k\={a}\'{s}a the qualities are---sound, number, dimension,
separateness, conjunction and disjunction. (VS 7.1.22)\\
Thus, then, being endowed with qualities, and not being located
in anything else, it is regarded as a substance. And in as much as
it has no cause, either homogeneous or heterogeneous, it is eternal.
(VS 2.1.18)\\
Time is the cause of the [relative] notions of ``priority,''
``posteriority,'' or ``simultaneity'' and ``succession,'' and
of ``late'' and ``soon.''
In as much as there is no other cause or basis for these
notions, as appearing with regard to these objects,---notions
which differ in character from all notions described before,---we
conclude ``time'' to be the basis of these. (VS 2.2.6)\\
Time is the cause or basis of the production,
persistence and destruction (or cessation) of all
produced things; as all these are spoken of in terms of time... (VS 2.2.9)\\
Though from the uniformity of the distinguishing character of time,
time is directly by itself, one only, yet, it is indirectly,
or figuratively, spoken of as manifold, on account of the
diversity among the conditions afforded by the production,
persistence and cessation of all produced things...

Space is the cause of the notions of east, west, 
below and above, and so on,
with regard to one material object considered with reference
to another material object as the starting point or limit.
Specially so, as there is no other cause for these notions.
(VS .2.12; 2.1.31; 7.1.24; 7.2.22)\footnotemark

\end{quote}

\subsection*{The nature of sound}

The underlying physical ideas of our systems are 
presented well in the discussion of sound. According to Pra\'{s}astap\={a}da:

\begin{quote}
Sound is the quality of \={a}k\={a}\'{s}a,
perceptible by the auditory organ.
It is momentary.
It can be produced by contact,
by disjunction, or by another sound.
There are two kinds of sound:
{\it var\d{n}a} (syllables) and {\it dhvani}.
The production of the syllables is a result of the contact
of the internal organ and self when influenced by memory.
First, one desires to produce the sound and then makes an effort.
The moving air strikes the throat, producing a contact with
the \={a}k\={a}\'{s}a, and resulting in the sound.
Sounds are always produced in a series, like a series of
ripples in water and when these waves reach the ear we hear them.\footnotemark
\end{quote}

Sound energy is viewed as a wave.
The waves impinge on the hearing organ and are recognized 
through associations.
Pra\'{s}astap\={a}da's dhvani is considered
to be noise.
But it appears that its role is similar to the
dhvani
defined by \={A}nandavardhana and
Abhinavagupta as the power of suggestion in its purest form
that plays a significant part in the recall of the
conscious and unconscious
associations.

\subsection*{Evolution}
With the background of the bandhu between the 
outer and the inner in mind, it is clear that the
evolution of the tattvas can also be viewed as
an evolution of the universe.
Buddhi or mahat
arises before space and matter.
This presumes that with buddhi also emerges the
cognition of time.
And further, that space and matter, which constitute the
physical universe, are contingent on the existence of
intelligence.
The working of the nature's intelligence is seen
as soon as the notions of prior and posterior,
related to the change associated with a
physical process,
become real.

The S\={a}\d{m}khya system also presupposes a universe which
comes into being and then is absorbed back in the ground-stuff
of reality.
This is what we see in the Pur\={a}\d{n}ic cyclic universe also.
Within each cycle, a gradual development of intelligent life is
assumed.
It is postulated that the plants arose first, followed by
animals of various kinds, and lastly by man.
Such a creation and destruction may be viewed to be
taking place at various levels, including the psychological level
related to the creation and destruction of thoughts.

\section{Analysis, causality}

The choice of the basic categories
in both S\={a}\d{m}khya and Vai\'{s}e\d{s}ika is dictated by
considerations of
economy.
This parallels a similar emphasis on economy in the
Indian grammatical tradition.
The fundamental bandhu
between language, thought and empirical reality
make it possible to analyze the processes of nature.

The S\={a}\d{m}khya
K\={a}rik\={a}s
present the question of {\it pram\={a}\d{n}a},
the method of validation, thus:
\begin{quote}
Perception, inference, and reliable authority are
considered the three means for this purpose.
Perception is the selective ascertainment of
particular sense-objects.
Inference, which is of three kinds, depends upon a
characteristic mark and that which bears that mark
(association). Reliable authority is
trustworthy verbal testimony.
The understanding of things beyond the senses is
inferred by analogy.\footnotemark
\end{quote}

The Vai\'{s}e\d{s}ika S\={u}tras also
clearly present the principle of cause ({\it k\={a}ra\d{n}a})
and effect ({\it k\={a}rya}).\footnotemark
Pra\'{s}astap\={a}da describes time and space as
{\it nimittak\={a}ra\d{n}a}, efficient cause, for all phenomena.
This indicates position in space and change in time are fundamental
to all reality.

Causality is expressed in S\={a}\d{m}khya as
{\it satk\={a}rya}, ``the doctrine of the
existence of the effect (in the cause).''
\begin{quote}
The effect exists due to: (a) the non-productivity of 
non-being; (b) the need for an appropriate material cause;
(c) impossibility of all things coming from all things;
(d) things producing only according to their nature;
(e) the nature of the cause.\footnotemark
\end{quote}

There is no {\it ex nihilo} creation in the
S\={a}\d{m}khya but only a progressive 
manifestation.

The gu\d{n}as provide the necessary ingredient
for the universe (be it physical or psychological)
to evolve.
They make it possible to distinguish between
the prior and the posterior.
The action of gu\d{n}as is 
essential to the definition of time and to the
workings of causality.

But gu\d{n}as are really not objective constituents of
nature.
Rather, they represent a relative property.
This is explained most clearly in {\it Gau\d{d}ap\={a}dabh\={a}\d{s}ya} in the
relativity inherent in ``the beautiful and virtuous woman who is a source
of delight but cause of pain to her co-wives and of delusion in the
passionate.''\footnotemark
In physical terms, one may speak of a separation between two
extremes by activity in the middle.
Or, the gu\d{n}as may be viewed as the potential whose gradients set
up the process of ceaseless change. 
The activity in the middle,
characterized by rajas, separates the
two poles of puru\d{s}a and undifferentiated prak\d{r}ti, or those
of sattva and tamas.

\section{How does the mind make sense?}

The observer has become a part of physics since the
advent of relativity and quantum mechanics;
the observer also plays a central role in Indian philosophical systems.
The question of observation in
S\={a}\d{m}khya and Vai\'{s}e\d{s}ika 
is considered at two levels:
at the level of the mind, which is seen as an
instrument; and at the level of the awareness ground-stuff, puru\d{s}a.

The S\={a}\d{m}khya model of the mind was shown in
Figure 2.
In it intellect (buddhi), self-consciousness (aha\.{m}k\={a}ra) and
mind (manas) are the three inner instruments that
process the sense impressions.

\begin{quote}

Since the buddhi together with the other internal
organs
(aha\.{m}k\={a}ra and manas) comprehends every object;
therefore, the three-fold instrument is the doorkeeper and
the remaining (ten) are the doors.\footnotemark
\end{quote}

Memory is seen to arise due to associations and the traces
let by past cognitions; this involves a contact between the
self and the internal organ.
The traces are stored by repetitions and by selective interest
in the objects of the past cognitions.
A recalled memory may become the cause of recollection of a part of
the previous cognition, desire or
aversion, and of further association of ideas.\footnotemark

Ordinary language is limited in its capacity to describe
all nature, likewise memories are inadequate in their remembrance of
the past.
But \={a}tman, by virtue of its linkages with brahman, does have
access to the hidden memories.
This means that a part of the mind is unconscious, inaccessible to
the empirical self.

Pra\'{s}astap\={a}da calls memory as a form of true knowledge
({\it vidy\={a}}) but does not count it as a pram\={a}\d{n}a.
The objection to memory being considered as true knowledge is that
it is just a trace.
A memory does not represent an object completely;
it leaves out some of the properties previously present and adds others
that were not initially there.
In other words, memories are reconstructions of reality. 

Cognition cannot be taken to arise out of the sense-organs.

\begin{quote}
These (organs, namely, 
aha\.{m}k\={a}ra, manas and the ten senses) which are
different from one another and which are distinct 
specifications of the gu\d{n}as, present the whole
to the buddhi, illuminating it for the puru\d{s}a like
a lamp.\footnotemark
\end{quote}

The question of the seat of intelligence
is analyzed::

\begin{quote}

In the cognitions of sound, etc, we infer a ``cognizer.''
This character cannot belong to the body, or to the
sense-organs, or to the mind; because all these are
unintelligent or unconscious.
Consciousness cannot belong to the body, as it is a material
product, like the jar; and also as no consciousness is found
in dead bodies.\\

Nor can consciousness belong to the sense-organs; because these
are mere instruments, and also because we have remembrances of objects
even after the sense-organ has been destroyed, and even when the
object is not in contact with the organ.\\

Nor can it belong to the mind; because if the mind be regarded
as functioning independently of the sense organs, then we would have
perception and remembrance simultaneously presenting themselves;
and because the mind itself is a mere instrument.\\

And thus the only thing to which consciousness could belong
is the self, which thus is cognized by this consciousness.\\

As from the motion of the chariot we infer the existence of an
intelligent guiding agent in the shape of the charioteer, so also
we infer an intelligent guiding agent for the body, from the activity
appearing in the body, which have the
capacity of acquiring the desirable and avoiding the
undesirable.\footnotemark
\end{quote}

Coming to the question of puru\d{s}a, it is stated 
{\it na prak\d{r}tir na vik\d{r}ti\d{h} puru\d{s}a\d{h},}
that it is 
neither {\it prak\d{r}ti} (creative) not {\it vik\d{r}ti} (created).\footnotemark
Puru\d{s}a transcends vyakta and avyakta, it is discriminating,
subjective, specific, conscious and non-productive.\footnotemark
Puru\d{s}a is
a witness, free, indifferent, watchful, and inactive.\footnotemark

The puru\d{s}a, in this characterization, does not interfere
with prak\d{r}ti and its manifestations. It is
transcendent and completely free ({\it kaivalya}).

What are the reasons that puru\d{s}a must exist?
\begin{quote}

{\it sa\d{m}gh\={a}tapar\={a}rthatv\={a}t\\
trigu\d{n}\={a}diviparyay\={a}d adhi\d{s}\d{t}h\={a}n\={a}t,\\
puru\d{s}o'sti bhokt\d{r}bh\={a}v\={a}t\\
kaivaly\={a}rtha\d{m} prav\d{r}tte\'{s} ca.}\\

The puru\d{s}a exists because aggregations exist for another;
because there must be the opposite to the three gu\d{n}as;
because there must be superintending power;
because there must be an enjoyer;
because there is activity for the sake of freedom.\footnotemark

\end{quote}
 
We see that this conception of the ``enjoyer'' or ``observer''
parallels the manner in which the observer enters the picture
in modern physics.\footnotemark
The physical laws are immutable; nevertheless, the universe 
appears to require that observers be present.

There is also the paradox that while corresponding to prak\d{r}ti
there exists a single puru\d{s}a, or a single root consciousness, in reality
there are many observers.

\begin{quote}
{\it jananamara\d{n}akara\d{n}\={a}n\={a}\d{m}\\
pratiniyam\={a}d ayugapatprav\d{r}tte\'{s} ca,\\
puru\d{s}abahutva\d{m} siddha\d{m}\\
traigu\d{n}yaviparyay\={a}c cai'va.}\\

The plurality of puru\d{s}as arises from: the diversity
of births, deaths, and faculties; actions or functions at 
different times; difference in the proportion of gu\d{n}as
in different individuals.\footnotemark
\end{quote}

The proximity between prak\d{r}ti and
puru\d{s}a makes it appear that the unconscious is endowed
with awareness.\footnotemark

In other words, the language of the k\={a}rik\={a}s does acknowledge
with great clarity, and in a manner perfectly consistent with 
modern insights, that the question of consciousness represents
a paradox.
The mind is taken to operate in a causal fashion, just as
the physical world does.
The sensory input is transformed by the associations
of different kinds that lie in the memory and
the predispositions (as determined by the gu\d{n}as)
to reach judgments.

\section{Qualities, motions, universals}

Ka\d{n}\={a}da lists seventeen qualities and says there
are more.
Candramati,
in {\it Da\'{s}apad\={a}rtha\'{s}\={a}stra},
adds the following seven to this list:
mass ({\it gurutva}), fluidity ({\it dravatva}),
viscidity ({\it sneha}), disposition ({\it sa\d{m}sk\={a}ra}),
merit ({\it dharma}), demerit
({\it adharma}), and sound ({\it \'{s}abda}).

Mass inheres in earth and water
and causes a substances to fall down.
Fluidity inheres in earth, water and fire and causes 
the flowing of a substance.
Viscidity inheres in water and causes coherence with a
substance such as earth.
Disposition can either be physical, in relation to a motion,
or mental.
Merit and demerit are psychological qualities related to
pleasure and pain.
Merit is of two kinds, viz., activity ({\it prav\d{r}tti})
and inactivity ({\it niv\d{r}tti}).

In physical terms, four states of matter are described: \={a}k\={a}\'{s}a
or ether,
which is non-atomic and, therefore, by itself represents vacuum;
gas, as in tejas; liquid, as in water;
and
solid, as in earth.
Since the aggregate substances have size, the question of
the manner in which their qualities inhere arises.

A distinction was made between qualities which pervade
their loci and those which do not.
Candramati 
lists the following
as locus-pervading:

\begin{tabular}{rrrrr}

color & taste & smell\\
touch & number & dimension \\
separateness & farness & nearness \\
contact & disjunction & fluidity \\
viscosity & weight & velocity \\

\end{tabular}

These are
the ones of significance for physical objects.
Sometimes, a few additional qualities are said to be
locus-pervading.

Pra\'{s}astap\={a}da describes qualities
related to objects somewhat differently than Candramati.
He offers
weight, fluidity, viscidity and 
sa\d{m}sk\={a}ra (disposition); this last quality is
further subdivided into
inertia ({\it vega}), elasticity ({\it sthitisth\={a}paka}), and 
trace ({\it bh\={a}van\={a}}).\footnotemark

Fluidity is of two varieties: natural and instrumental.
It is a natural quality of water and an instrumental
quality of earth and fire.
When water freezes into ice, the natural fluidity of water is
seen to be counteracted by the fire of the sky, so that the atoms
combine to form a solid.
Water, earth, and fire all have fluidity.
However, water's fluidity is held to be primary,
while that of the other two substances is secondary.
Viscidity is responsible for cohesion and smoothness.

Ka\d{n}\={a}da defines motion into five varieties:
ejection
({\it utk\d{s}epa\d{n}a}), attraction
({\it avak\d{s}epa\d{n}a}), contraction ({\it \={a}ku\~{n}cana}), expansion
({\it prasara\d{n}a}), and composite movement ({\it gamana}).\footnotemark
In the case of gamana there is contact with
points of space in various directions, or there are many
loci.\footnotemark
Motion by gravity is discussed.
``Weight causes falling; it is imperceptible and known
by inference.''\footnotemark
Motion is produced by mass, which is the same
as a motion due to gravitational attraction.\footnotemark

Inertia is the quality of a moving object which is responsible
for its continuing in its motion.
The Vai\'{s}e\d{s}ika position is that inertia is countered
by other forces, leading to energy
loss, which is why the moving object slowly
loses its speed.

That motion cannot take place instantaneously, was well understood.
Vyoma\'{s}iva in his Vyomavat\={\i} speaks of how a motion has several
parts that will take increments of time.
Likewise, motions produced in cooking will take time to produce
the new quality associated with the process, where
time, in this context, is equivalent to energy.
This is a statement of the empirical fact that a minimum
energy needs to be expended before a state change occurs.
With water the temperature must reach the boiling point before
steam will be obtained.
This observation expresses an understanding of the quantum
effect in daily processes.

It is stated that there are two kinds of universals:
higher and lower.\footnotemark
The higher universal here is Being, which encompasses everything.
Lower universals exclude as well as include.
This means that the universals could be defined  in a
hierarchical fashion.
The higher universal is akin to a superposition of all
possibilities and so it anticipates the essence of
the quantum theory.

\section{Cosmology, astronomy}

We now consider how the ideas of S\={a}\d{m}khya and
Vai\'{s}e\d{s}ika are intertwined with the development of
Indian science.
Since S\={a}\d{m}khya, in one of its many forms, has
been a part of Indian thinking going back to the remotest
times, one may be certain that it played an
important role.
This is most easy to see for
astronomy for which the
extant texts provide enough information in terms
of layers of material, and thereby allow us to see a gradual
development of various ideas.
This evolution of astronomy may be taken to be
a prototype for the development of other sciences.

Ideal forms play a role in
Vai\'{s}e\d{s}ika.
For example, sphericity ({\it p\={a}rima\d{n}\d{d}alya})
is considered a basic shape.
Candramati speaks of two kinds of sphericity:
when it is minute, it resides in an atom,
and when it is absolutely large (infinite),
it resides in \={a}k\={a}\'{s}a, time, place, and self.
In between the very large (cosmos) and the very small
(atom) are the objects of the observable universe
which will not conform to the ideal shape.
So in astronomy, which represents
this middle ground, one must consider deviations from
spherical or circular shapes and orbits.

Since only the cosmos as a whole may be considered
to be perfect, space as a dravya will not
have any absolute properties.
This reasoning sets Indian physical science apart from
the tradition of Greek science which took space
to be absolute and the
observer on the earth to have
a privileged position.
In Indian physics,
space and time are considered to be relative.

Considering Indian astronomy, it should  be noted that its
understanding is undergoing
a major shift. More than a hundred years ago, it was believed\footnotemark that
the Indians were the originators of many of the notions that
led to the Greek astronomical flowering.
This view slowly lost support and then it was believed that
Indian astronomy was essentially derivative and it
owed all its basic ideas to the Babylonians and the Greeks.
It was even claimed that there was no tradition of
reliable observational astronomy in India.

Billard,\footnotemark using 
statistical analysis of the parameters used
in the many Siddh\={a}ntas, showed that
these texts were based on precise observations and
so the theory that there was no observational tradition in India was
wrong.
Seidenberg showed that the altars of the Br\={a}hma\d{n}as already
knew considerable geometry.
He saw the development of the mathematical
ideas going through the sequence of equivalence by number followed
by an equivalence by area.\footnotemark
Further work showed that these altars represented
astronomical knowledge.
Since then it has been found that the Vedic books are
according to an astronomical plan.\footnotemark
The texts themselves mimic the tripartite
connections of nature!

\subsection*{On the non-uniform motion in the sky}

We first consider the sun.
With respect to an observer on the earth, the sun has two 
motions.
First, is the daily motion across the sky. Second, is the 
shifting of the rising and setting directions. 
It is this second motion which defines the seasons.
Its two extreme points are the solstices, and the points where the
sun's orbit crosses the equator or when the
nights equal the days are the equinoxes.

The Aitareya Br\={a}hma\d{n}a describes how the sun reaches the highest point
on the day called vi\d{s}uvant and how it stays still for a
total of 21 days with the vi\d{s}uvant being the middle day of
this period.
In the Pa\~{n}cavi\d{m}\'{s}a Br. several year-long rites
are described where the vi\d{s}uvant day is preceded and followed
by three-day periods called svaras\={a}man days. This suggests
that the sun was now taken to be more or less still in the heavens
for a total period of 7 days.
So it was clearly understood that the shifting of the rising and
the setting directions had an irregular motion.

The year-long rites list a total of 180 days before the solstice and
another 180 days following the solstice. 
Since this is reckoning by
solar days, it is not clear stated how
the remaining 4 or 5 days of the year were
assigned.
But this can be easily inferred.

The two basic days in this count are the
vi\d{s}uvant (summer solstice) and the mah\={a}vrata day
(winter solstice)
which precedes it by 181 days in the above counts.
Therefore, even though the count of the latter part of the
year stops with an additional 180 days, it is clear that
one needs another 4 or 5 days to reach the
mah\={a}vrata day in the winter.
This establishes that the division of the year was in the
two halves of 181 and 184 or 185 days.
Corroboration of this is 
suggested by evidence related to an
altar design from \'{S}atapatha Br\={a}hma\d{n}a  as
shown in Fig 4.
This figure shows that the four quarters of the year were not
taken to be equal.\footnotemark

\begin{figure}
\hspace*{0.2in}\centering{
\psfig{file=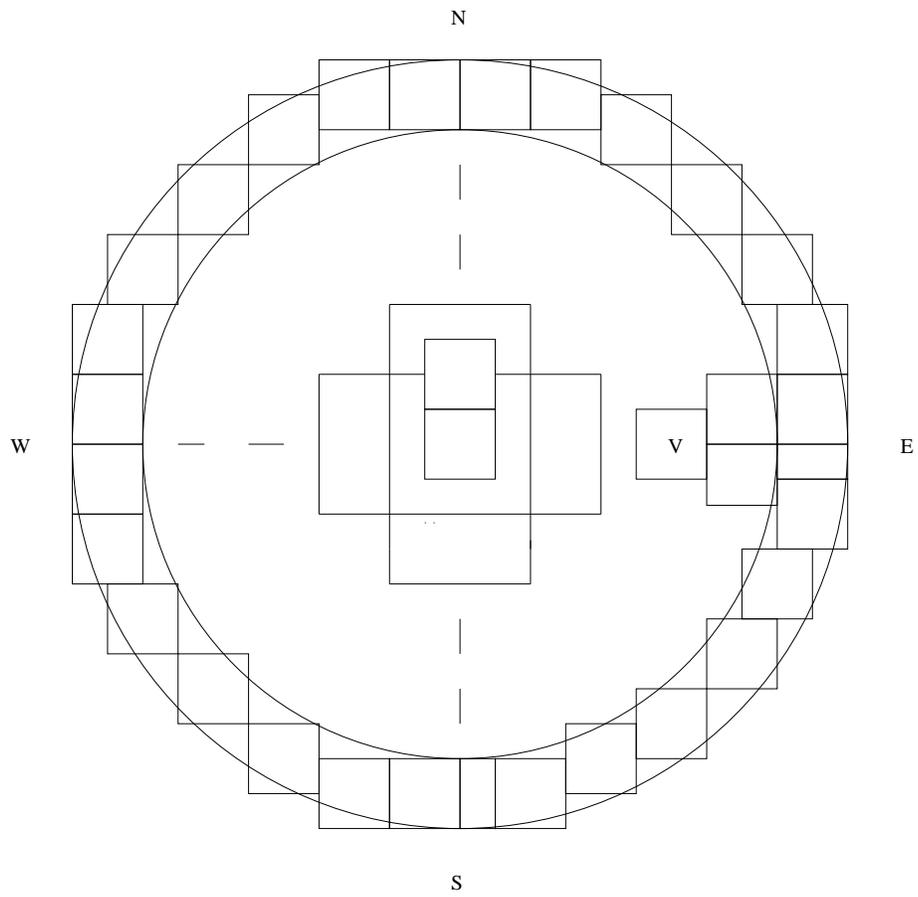,width=12cm}}
\caption{The non-uniform circuit of the Sun}
\end{figure}

Likewise, the motions of the planets were known to be
non-uniform.
The ideal orbits were considered to be circular.
But the actual motion deviated from the ideal,
represented in terms of the struggle betweeen
the devas and the asuras of the Vedic
mythology.\footnotemark

By the time of the Siddh\={a}ntas,
the planet orbits were represented with respect to
the sun.
Not only did \={A}ryabha\d{t}a (c. 500 C.E.) believe that the earth rotates,
but there are glimmerings in his system (and other similar
Indian systems) of a possible underlying theory in which the
earth (and the planets) orbits the sun, rather than the sun
orbiting the earth.
The evidence is that the period provided is 
the \'{s}\={\i}ghrocca, which is the time taken by the planet
to orbit the sun
relative to the sun.
For the outer planets this is not significant: both earth and
sun are inside their orbits and so the time taken to
go round the earth and the time taken to go round the sun
are the same.
But this becomes
significant for the inner planets.

\subsection*{The motion of the earth}

Only an ideal body will be at complete rest in the
Vai\'{s}e\d{s}ika system.
So it is not surprising to see
\={A}ryabha\d{t}a take the earth to rotate
on its axis.
It appears that the rotation of the earth
is inherent in the notion that the sun never sets that
we find in the Aitareya Br\={a}hma\d{n}a:

\begin{quote}
The [sun] never really sets or rises.
In that they think of him ``He is setting,'' having reached the
end of the day, he inverts himself; thus he makes evening below,
day above. Again in that they think of him ``He is rising in the
morning,'' having reached the end of the night he inverts himself;
thus he makes day below, night above. He never sets; indeed he
never sets.\footnotemark
\end{quote}

One way to visualize it is to see the universe as the hollow of
a sphere so that the inversion of the sun now shines the light
on the world above ours. But this is impossible
since the sun does move across the sky during the day and
if the sun doesn't set or rise it doesn't move either.
Clearly, the idea of ``inversion'' denotes nothing but a movement
of the earth.

By our study of the early Vedic sources, we are are now 
able to understand the stages of the
development of the earliest astronomy.
After the \d{R}gvedic stage comes the period 
of the Br\={a}hma\d{n}as.
This is followed by Lagadha's astronomy.
The last stage is early Siddh\={a}ntic and early Pur\={a}\d{n}ic
astronomy.

Ancient Indian astronomy may be described in
terms of three broad stages 
which mirror the development of the corresponding
philosophical ideas.

\begin{itemize}

\item {\it Stage 1}.
This is primarily \d{R}gvedic astronomy.
Here we speak of
the motion of the sun and the moon, nak\d{s}atras, and the planets
and their
orbits are not according to the ideal
cyclic motions.
We are not certain when this period began
but we have many references of astronomical events in the 
mythology, like the destruction of the sacrifice of Dak\d{s}a
by \'{S}iva, which indicates the era of the fourth
millennium B.C.E.,\footnotemark
but note that this story belongs to a later stratum of the
Vedic myths.
The ritual of this period was done according to the
Ved\={a}\.{n}ga Jyoti\d{s}a of Lagadha (c. 1300 B.C.E.).
Being the standard manual for determination of the Vedic rites,
Lagadha's work must have served as a ``living'' book which
is why the
language of the extant text shows later linguistic usage.
The objective of the
Ved\={a}\.{n}ga Jyoti\d{s}a
is to do the astronomical calculations for
the daily rites.
The calculations use forms that may be unrelated to
observational processes as in the case of the  mean 
positions of the sun and the moon.
The day is defined with respect to the risings of the sun,
the stars, and the moon.
This indicates that the idea of
relativity with respect to time processes was well understood.
The orbits of the sun and the moon are considered with respect 
to their mean positions, suggesting the non-uniform nature of their
motions was well known.

\item {\it Stage 2.}
This is the astronomy of the Br\={a}hma\d{n}as associated with
names like those of
Y\={a}j\~{n}avalkya and \'{S}\={a}\d{n}\d{d}ilya.
Although the rites of the 
Br\={a}hma\d{n}as appear to be very ancient,
the texts appear to belong to the
second millennium B.C.E.
Their astronomy is represented by means of geometric altars
and deals with the non-uniform
motion of the sun and the moon and intercalation for the lunar year.
There is also the beginnings of an understanding of
universal attraction in terms of 
``strings of wind joined to the sun.''
This astronomy corresponds to the S\={a}\d{m}khya of
the 
B\d{r}had\={a}ra\d{n}yaka Upani\d{s}ad.

\item {\it Stage 3.}
This concerns mainly with
the early Siddh\={a}ntic and Pur\={a}\d{n}ic periods.
Here our main sources are the
\'{S}ulbas\={u}tras, the Mah\={a}bh\={a}rata, the
early Pur\={a}\d{n}as, S\={u}ryasiddh\d{a}nta and other texts.
This stage saw the
development of the \'{s}\={\i}ghrocca and mandocca cycles and
the concept of the kalpa, the large period associated with creation
at the cosmic level.

\end{itemize}

At the end of these stages stands the classical 
Siddh\={a}ntic period inaugurated by
\={A}ryabha\d{t}a.
The concepts of the 
\'{s}\={\i}ghrocca and mandocca cycles indicate that the motion of
the planets was taken to be fundamentally around the sun, which, in
turn, was taken to go around the earth.
We can see the development of these ideas as an explication of
the notion of a non-ideal motion in terms of several
stages of underlying ideal motions.
This is analogous to how the ideal shapes of the atoms,
when combined, lead to the non-ideal shapes of the
gross elements.

{\it S\={u}rya Siddh\={a}nta} describes a ``mechanistic''
model for the planetary motions which is like the
mechanistic physical models of the Vai\'{s}e\d{s}ika:

\begin{quote}
Forms of time, of invisible shape, stationed in the zodaic,
called the \'{s}\={\i}ghrocca, mandocca, and node (p\={a}ta),
are causes of the motion of the planets.
The planets, attached to these points by cords of air,
are drawn away by them, with the right and left hand, forward or
backward, according to nearness, toward their own place.
A wind, called {\it pravaha}, impels them toward their own
uccas, being drawn away forward and backward.\footnotemark
\end{quote}

The antecedents of this system can be seen in the earlier texts.
\d{R}gveda speaks of the stars of the Ursa Major
(the Seven Sages) having ropes of wind,
({\it munayo v\={a}ta ra\'{s}an\={a}\d{h}}).\footnotemark
\'{S}atapatha Br. describes the sun as 
{\it pu\d{s}karam\={a}dityo}, ``the lotus of the sky.''
It also says:

\begin{quote}
{\it tadas\={a}v\={a}ditya im\={a}\d{m}lok\={a}nts\={u}tre
sam\={a}vayate, tadyattats\={u}tra\d{m} v\={a}yu\d{h}..}

The sun strings these worlds [the earth, the planets,
the atmosphere] to himself on a thread.
This thread is the same as the wind...\footnotemark
\end{quote}

This suggests a central role to the sun in defining the motions
of the planets and ideas such as these must have ultimately
led to the theory of the \'{s}\={\i}ghrocca and the
mandocca cycles.

The sun's central role implies that
the basic functioning
of gravitation was understood.
This action was visualized in terms of ``ropes of wind,''
which, in modern terminology, would be called
a field.

\subsection*{Relativity of time and space}

To summarize, the
first descriptions were non-uniform motions of ``mean''
objects. 
Later models shift the centre from the earth, first by
considering that earth spins on its axis, and then representing
the non-circular motion of the planets with respect to the sun.

The parallel speculative thought in the Pur\={a}\d{n}as
takes space and time to be relative in a variety of ways.
Time can flow at different rates for different observers.
Time and space are not absolute.
There exist countless universes with their own
Brahm\={a}, Vi\d{s}\d{n}u, and Mahe\d{s}a.

To appreciate the background for this thought, consider that
in Vai\'{s}e\d{s}ika the universal is taken to be timeless and
ubiquitous.
Whatever can be defined with respect to space and time cannot
be a universal.
The processes that mark the passage of time on an object would
thus be relative.
It is only the universals which are of the highest form,
i.e. true for all time and space,
that are absolute.
And the only such universal is the Being.

These ideas are elaborated in the Pur\={a}\d{n}as, the
\={A}gamic and the T\={a}ntric literature, and in
books like the Yoga-V\={a}si\d{s}\d{t}ha.

\section{Concluding remarks}

We have shown that the physical concepts underlying
S\={a}\d{m}khya and Vai\'{s}e\d{s}ika represent a 
sophisticated materialist framework for the laws of nature.
This physics was based on general observations on the
various physical processes.
Since an element of the two philosophical systems was
metaphysical, the reasoning was often validated based
on psychological arguments.
Both
systems emphasized causality and so were
capable of elucidating nature's laws.
The basic categories are ideals and 
the modifications of these ideas provide endless
structure.

There is also a
complementarity between S\={a}\d{m}khya and
Vai\'{s}e\d{s}ika.
By considering the evolution of tattvas, S\={a}\d{m}khya
emphasizes genesis both at the cosmic as well as
the psychological levels.
More details related to the constitution of the physical world
are provided by 
Vai\'{s}e\d{s}ika.
These structures are paralleled in Indian
grammatical philosophy with 
production based on a small set of axioms.

There is also a recognition that new enumerative categories
are needed in the characterization of empirical world.
It is recognized
that the description of the physical world requires 
categories that go beyond the basic 25 of the S\={a}\d{m}khya system.
Some of them are described in Vai\d{s}e\d{s}ika,
but there the emphasis is on atoms and their
mutual relationships.
For example, new categories are necessary to characterize the motion of
planets.
Driven by the requirement of
reconciling the cyclic ideal motions of the planets to the
actual ones, more complex orbits were introduced.
This complexity was seen as being engendered by
the workings of
gravity-like forces.

Speaking of one of these philosophies, the historian of
thought Karl Potter says:

\begin{quote}
Ny\={a}ya-Vai\'{s}e\d{s}ika offers one of the most vigorous
efforts at the construction of a substantialist, realist ontology
that the world has ever seen.
It provides an extended critique of event-ontologies and
idealist metaphysics.
It starts from a unique basis for ontology that incorporates
several of the most recent Western insights into the question
of how to defend realism most successfully.
This ontology is ``Platonistic'' (it admits
repeatable properties as Plato's did),
realistic (it builds the world from ``timeless''
individuals as well as spatio-temporal points or events),
but neither exclusively physicalistic nor phenomenalistic
(it admits as basic individuals entities both directly known
and inferred from scientific investigations).
Though the system has many quaint and archaic features
from a modern point of view, as a philosophical base
for accommodating scientific insights it has advantages:
its authors developed an atomic theory, came to treat numbers
very much in the spirit of modern mathematics, argued for a wave
theory of sound transmission, and adapted an empiricist
view of causality to their own uses.\footnotemark
\end{quote}

In reality, the scope of S\={a}\d{m}khya and Vai\'{s}e\d{s}ika is even
greater than this, because they reconcile the
observer to the frame of a materialist physics,
leading to subtle insights that have been validated by 
modern physics.
Consider, for example, the notion that one may take the
tanm\={a}tras to be composed of bh\={u}t\={a}di or the
other way round.
The tanm\={a}tras are an abstract potential whereas 
bh\={u}t\={a}di are the elementary atoms
which is somewhat like the quantum wavefunction and material particles.
S\={a}\d{m}khya, where the observer is central, considers
tanm\={a}tras to emerge first.
On the other hand, Vai\'{s}e\d{s}ika. with its focus on atoms and
their combinations, does not speak of tanm\={a}tras although
some of the gu\d{n}as are like the tanm\={a}tras.
In other words, we have something akin to the
concept of wave-particle duality 
of quantum physics.

The assumption that all observed world emerges out of
prak\d{r}ti implies that the material substratum of
all substances is the same.
The qualities of Vai\'{s}e\d{s}ika emerge as material atoms
combine in different ways.
These emergent properties are not limited to inanimate matter
but also to the instruments of cognition where
actual cognition requires the self to be
the activating agent.

This paper presents only two of the many currents of the
Indian physical thought. One needs also to consider texts
on architecture, astronomy  as well as the traditions
related to crafts and military science for additional
insight.

\section*{Abbreviations}
\begin{tabular}{ll}
AB &	 Aitareya Br\={a}hma\d{n}a\\
AV &	 Atharvaveda\\
BU &	 B\d{r}had\={a}ra\d{n}yaka Upani\d{s}ad\\
CU &	 Ch\={a}ndogya Upani\d{s}ad\\
KBU &    Kau\d{s}\={\i}taki Br\={a}hma\d{n}a Upani\d{s}ad\\
PP &	 Pad\={a}rthadharmasa\d{m}graha of Pra\'{s}astap\={a}da\\
RV &	 \d{R}gveda\\
\'{S}B &	 \'{S}atapatha Br\={a}hma\d{n}a\\
SK &	S\={a}\d{m}khya K\={a}rik\={a} \\
\'{S}U & \'{S}vet\={a}\'{s}vatara Upani\d{s}ad\\
TU &	 Taittir\={\i}ya Upani\d{s}ad\\
VS &	 Vai\'{s}e\d{s}ika S\={u}tra\\
\end{tabular}


\section*{Notes}

\begin{enumerate}

\item SK 2.

\item Heisenberg [5], Kak [9,14,16-18,21].

\item RV 1.164.45.

\item See, for example, Kak [7,8,10-15,20].

\item RV 4.23; 10.85; 10.190.

\item AV 10.

\item RV 2.25; 10.121.

\item RV 10.129.4-5.

\item BU 4.5.12-13.

\item BU 1.4.1.

\item KBU 3.5.

\item CU 7.25.1.

\item CU 6.2-5.

\item \'{S}U 4.5.

\item TU 2.1.

\item CU 6.2.1-2.

\item TU 3.2-6.

\item See, for example, Dasgupta [3],
Matilal [28], Potter [29], Hulin [6],
Larson [26], Larson and Bhattacharya [27].

\item VS 2.1
\item Seal [31], pages 37-38.

\item Seal [31], pages 38-39.

\item PP 5.41-3.

\item PP 137.

\item SK 4-7.
\item SK 7.
\item SK 9.

\item Gau\d{d}ap\={a}dabh\={a}\d{s}ya 9.

\item SK 35.

\item PP 121.

\item SK 36.

\item PP 5.44s; see also VS 3.1.19.

\item SK 3.

\item SK 11.

\item SK 19.

\item SK 17.

\item Heisenberg [5], Kak [9].

\item SK 18.

\item SK 20.

\item PP 129-133.

\item VS 10.

\item PP 143.

\item PP 129.

\item PP 149.

\item PP 154.

\item Burgess [2].

\item Billard [1].

\item Seidenberg [32].

\item Kak [15-20,22-25].

\item The figure is from \'{S}B 8.5; see also Kak [15].

\item Kak [15].

\item AB 4.18.

\item Kramrisch [25], pages 42-43.

\item S\={u}rya Siddh\={a}nta 2.1-5.

\item RV 10.136.2.

\item \'{S}B 8.7.3.10.

\item Potter [29], page 1.

\end{enumerate}

\section*{Bibliography}

\begin{description}
\item [1] R. Billard, {\it L'Astronomie Indienne.}
Ecole Francaise d'Extreme Orient, Paris, 1971.

\item [2] E. Burgess, {\it The S\={u}rya Siddh\={a}nta}.
Motilal Banarsidass, Delhi, 1989 (1860).

\item [3] S. Dasgupta, {\it A History of Indian Philosophy.}
Cambridge University Press, Cambridge, 1932.
 
\item [4] Dharampal, {\it Indian Science and Technology in the Eighteenth
Century.} Impex India, Delhi, 1971. 	

\item [5] W. Heisenberg, {\it Physics and Philosophy.}
Penguin, London, 1989.

\item [6] M. Hulin, {\it S\={a}\d{m}khya Literature}.
Otto Harrassowitz, Wiesbaden, 1978.
 
\item [7] S. Kak, 
``The astronomy of the age of geometric altars,"
{\em Quarterly Journal Royal Astronomical Society}, 36, 385-396, 1995.

\item  [8] S. Kak, 
``Knowledge of planets in the third millennium BC,"
{\em Quarterly Journal Royal Astronomical Society}, 37, 709-715, 1996.

\item [9] S. Kak, ``On the science of consciousness in ancient India,''
{\em Indian Journal of the History of Science,} 32, 105-120, 1997a.

\item [10] S. Kak, ``Archaeoastronomy and literature,''
{\it Current Science}, 73, 624-627, 1997b.

\item [11]
S. Kak, ``Speed of light and Puranic cosmology.''
ArXiv: physics/9804020.

\item [12]
S.Kak, ``Early theories on the distance to the sun.''
{\it Indian Journal of History of Science}, vol. 33, 1998, pp. 93-100.
ArXiv: physics/9804021.

\item [13]
 S. Kak, ``The solar numbers in Angkor Wat.''
{\it Indian Journal of History of Science}, vol. 34, 1999, pp. 117-126.
ArXiv: physics/9811040.

\item [14]
 S. Kak, ``Concepts of space, time, and consciousness in ancient India.''
ArXiv: physics/9903010.

\item [15]
 S. Kak, {\em The Astronomical Code of the \d{R}gveda}.
Munshiram Manoharlal, New Delhi, 2000.

\item [16]
 S. Kak, ``Birth and early development of 
Indian astronomy.'' In a book on {\it Astronomy Across Cultures:
The History of Non-Western Astronomy}, Helaine Selin (editor),
Kluwer Academic, Boston, 2000, pp. 303-340.
ArXiv: physics/0101063.  

\item [17]
 S. Kak,
`Astronomy and its role in Vedic culture.'' In
{\it Science and Civilization in India, Vol. 1:
The Dawn of Indian Civilization}, Part 1, edited by G.C. Pande,
ICPR/Centre for Studies in Civilizations, New Delhi, 2000, pp. 507-524.

\item [18]
 S. Kak,
``Physical concepts in Samkhya and Vaisesika.'' In
{\it Science and Civilization in India, Vol. 1, Part 2, Life, Thought and
Culture in India (from c 600 BC to c AD 300)}, edited by G.C. Pande,
ICPR/Centre for Studies in Civilizations, New Delhi, 2001, pp. 413-437.

\item [19]
S. Kak, On Aryabhata's planetary constants. ArXiv: physics/0110029.

\item [20]
 S. Kak, {\it The Wishing Tree.} Munshiram Manoharlal, New Delhi,
2001.

\item [21]
S. Kak, ``The cyclic universe: some historical notes.''
ArXiv: physics/0207026

\item [22]
S. Kak, ``Greek and Indian cosmology: review of early history.''
ArXiv: physics/0303001.

\item [23]
S. Kak,
``Babylonian and Indian astronomy: early connections.''
ArXiv: physics/0301078.

\item [24]
S. Kak, ``Yajnavalkya and the origins of puranic cosmology.''
{\it Adyar Library Bulletin}, vol 65, pp. 145-156, 2001. Also in
ArXiv: physics/0101012.

\item [25] S. Kramrisch, {\it The Presence of \'{S}iva.}
Princeton University Press, Princeton (1981), pages 42-43.

\item [26] G.J. Larson, {\it Classical S\={a}\d{m}khya}.
Motilal Banarsidass, Delhi, 1979.

\item [27] G.J. Larson and 
R.S. Bhattacharya (ed.). {\it S\={a}\d{m}khya: A Dualist Tradition in 
Indian PhilosophY.}
Princeton University Press, Princeton, 1987.

\item [28] B.K. Matilal, {\it Ny\={a}ya-Vai\'{s}e\d{s}ika.}
Otto Harrassowitz, Wiesbaden, 1977.

\item [29] K.H. Potter (ed.). {\it Indian Metaphysics and Epistemology.}
Princeton University Press, Princeton, 1977.

\item [30] T.R.N. Rao and S. Kak (eds.). {\it Computing Science in Ancient
India.} Munshiram Manoharlal, New Delhi, 2000.

\item [31] B. Seal, {\it The Positive Sciences of the Hindus.}
Motilal Banarsidass, Delhi, 1985 (1915)

\item [32]  A. Seidenberg, ``The origin of mathematics,''
{\em Archive for History of Exact Sciences},
18, 301-342, 1978.

\end{description}

\end{document}